\documentclass[trackchanges,twocolumn]{aastex631}
\usepackage{etoolbox}          

\AtBeginEnvironment{figure}{\nolinenumbers}
\AtEndEnvironment{figure}{\linenumbers}
\AtBeginEnvironment{figure*}{\nolinenumbers}
\AtEndEnvironment{figure*}{\linenumbers}
\AtBeginEnvironment{table}{\nolinenumbers}
\AtEndEnvironment{table}{\linenumbers}
\AtBeginEnvironment{table*}{\nolinenumbers}
\AtEndEnvironment{table*}{\linenumbers}
\AtBeginEnvironment{deluxetable}{\nolinenumbers}
\AtEndEnvironment{deluxetable}{\linenumbers}
\AtBeginEnvironment{deluxetable*}{\nolinenumbers}
\AtEndEnvironment{deluxetable*}{\linenumbers}

\newcommand*\patchAmsMathEnvironmentForLineno[1]{%
  \expandafter\pretocmd\csname #1\endcsname{\linenomath}{}{}%
  \expandafter\pretocmd\csname end#1\endcsname{\endlinenomath}{}{}%
}
\newcommand*\patchBothAmsMathEnvironmentsForLineno[1]{%
  \patchAmsMathEnvironmentForLineno{#1}%
  \patchAmsMathEnvironmentForLineno{#1*}%
}
\AtBeginDocument{%
  \patchBothAmsMathEnvironmentsForLineno{equation}%
  \patchBothAmsMathEnvironmentsForLineno{align}%
  \patchBothAmsMathEnvironmentsForLineno{flalign}%
  \patchBothAmsMathEnvironmentsForLineno{alignat}%
  \patchBothAmsMathEnvironmentsForLineno{gather}%
  \patchBothAmsMathEnvironmentsForLineno{multline}%
}
\makeatletter
\let\frontmatter@title@above=\relax
\makeatother

\usepackage{subfig}
\usepackage{url}
\usepackage{amsmath}

\usepackage{ amssymb }
\usepackage{appendix}
\usepackage{float}
\usepackage{lipsum}

\shorttitle{The Effect of a Self-bound EOS on the Structure of Rotating Compact Stars}

\accepted{by ApJ, December 12, 2025}
\graphicspath{Figures/}
    \usepackage{graphicx}
    \usepackage{capt-of}
    \usepackage{lipsum}

\begin{document}

\correspondingauthor{}
\email{andreas.konstandinu@gmail.com}

\author[0000-0002-1072-7313]{Andreas Konstantinou}
\affiliation{Department of Physics, University of Cyprus, P.O. Box 20537, 1678 Nicosia, Cyprus}
\affiliation{Computation-based Science and Technology Research Center,
The Cyprus Institute, 20 Kavafi Str., Nicosia 2121, Cyprus}

\title{The Effect of a Self-bound Equation of State on the Structure of Rotating Compact Stars}

\begin{abstract}
This paper investigates how a self bound equation of state (EOS), which describes strange quark stars, affects the rotational properties of compact stars, focusing on deviations from universal relations governing gravitational mass and radius changes due to rotation. The analysis reveals significant deviations in stars with higher surface-to-center total energy-density ratios, $\frac{\epsilon_s}{\epsilon_c+c^2P_c}$, challenging the established universal relations.

For Newtonian stars, hydrostatic equilibrium ensures that the difference between the gravitational potential at the center, $\Phi_c$, and at the poles, $\Phi_p$, remains constant within sequences of rotating neutron stars characterized by the same central and polar specific enthalpy ($\Phi_c - \Phi_p = -h_c +h_p$). Combined with the scaling $\Phi \propto R_e^2$, where $R_e$ denotes the equatorial radius, this condition naturally leads to a quasi-universal behavior in the rotational change of radius within these sequences. Similarly, in general relativistic stars, the hydrostatic equilibrium maintains that $\Phi^{GR}_{p} - \Phi^{GR}_{c}$ remains unchanged within these sequences, where $\Phi^{GR}$ is one of the metric potentials.

Inspired by this theoretical framework, a toy model has been developed to capture the dependence of gravitational mass and radius deviations on the surface-to-central total energy density ratio. Subsequently, an improved set of empirical universal relations has been proposed, for accurately modeling rapidly rotating compact stars with self-bound EOSs.
\end{abstract}

\keywords{quark stars, neutron stars, Newtonian stars, relativistic stars, pulsars, stellar rotation, compact objects, millisecond pulsars}

\section{Introduction} \label{sec:intro}

Neutron stars (NSs), the compact remnants of supernova explosions, exhibit a wide range of fascinating and extreme physical phenomena because of their high densities and strong gravitational fields. Understanding the internal structure and properties of neutron stars requires a deep understanding of their equation of state (EOS), which describes how matter behaves at the immense densities found in these stars \citep{Lattimer_2004,Ozel:2016oaf,Baym:2017whm}.
In most EOS models describing hadronic matter, the pressure decreases smoothly to zero as the density tends to zero
\citep[e.g.][]{1971ApJ...170..299B,Haensel:2007yy,Douchin:2001sv}. Such EOSs are commonly referred to as \textit{normal} EOSs \citep{Lattimer:2001jsw}.

A distinct class of models describes \textit{self-bound} compact stars, most notably strange quark stars, whose matter is stabilized by the strong interaction \citep{WEBER2005193}. In these EOSs the pressure vanishes at a finite, non-zero energy density, and the stellar surface corresponds to a sharp discontinuity in density. The most well-established examples are strange quark stars, composed of deconfined up, down, and strange quarks, as proposed in the Bodmer–Witten–Terazawa hypothesis \citep{PhysRevD.4.1601,PhysRevD.30.272,1989JPSJ...58.3555T}. Other possibilities include strangeon-star models, in which matter is arranged into a gigantic three-flavor nucleus \citep{Lai:2017ney}. In the present work, however, the analysis is restricted to the standard strange quark star scenario.
A characteristic observational signature of self-bound configurations is the possibility of unusually small stellar radii, which has long been regarded as a potential ``smoking gun" for strange quark stars \citep{1986A&A...160..121H,1986ApJ...310..261A,Glendenning:1997wn}.

A separate and rapidly growing line of work concerns universal relations among macroscopic stellar quantities that are remarkably insensitive to the choice of EOS. Examples include the relationship between dimensionless moment of inertia and compactness \citep{1994ApJ...424..846Ravenhall}, the correlation between dimensionless gravitational binding energy and compactness \citep{2001ApJ...550..426Lattimer}, the relationship between the maximum mass of rotating and non-rotating neutron stars \citep{1996ApJ...456..300Lasota,2016MNRAS.459..646Breu}, and the ``I-Love-Q" relations, which link dimensionless moment of inertia, quadrupole moment, and the Love number \citep{2013Yagi,2017PhR...681....1Yagi}. 

Several studies examine how the presence of quark matter affects universal relations \citep{2013Yagi,Sham_2015,10.1093/mnras/stac1916,PhysRevD.95.101302,Bandyopadhyay_2018,Pretel_2024,PhysRevD.111.063026,59d4-4x9m}. For example, \citet{Doneva_2013} show a breakdown of the ``I-Love-Q'' relations for rapidly rotating strange stars.
Inspired by \cite{Doneva_2013}, this work investigates how strange quark star EOSs 
affect the universal relations that govern gravitational mass and radius changes due to rotation in sequences of compact stars with constant central energy density \citep{Konstantinou_2022}. The goal is to quantify deviations from the universal behavior seen in hadronic stars and to determine whether modified or entirely new relations are required for self-bound configurations.

Furthermore, it has been argued that the universal relation governing the change in the equatorial radius is linked to the fact that the gravitational potential difference between the center and poles remains constant in these sequences. This property has been used to create a toy model that will allow a better generalization of the universal relations in the self-bound regime. This constant potential quantity may also play a role in explaining why the equatorial compactness, defined as $M/R_e$, remains nearly constant along a sequence \citep{Konstantinou_2022}.

The structure of the paper is as follows. Section \ref{sec:compmeth} describes the methods employed to compute the structural properties of rotating strange quark stars using various EOSs. In Section \ref{sec:Results}, the computational results are presented, highlighting key deviations from established universal relations for hadronic compact stars. New empirical universal relations are introduced, which are accurate for strange quark star EOSs. Section \ref{sec:Discussion} explores the theoretical basis for these deviations, examining the conservation of certain potential quantities, such as the gravitational potential difference between the center and poles, and how these relate to hydrostatic equilibrium. This section also employs a toy model to explain the observed changes in gravitational mass and radius due to rotation. Finally, Section \ref{sec:conclusion} summarizes the key findings of this paper.

\section{Computational Methods}\label{sec:compmeth}
The \textit{Rapidly Rotating Neutron Star} code, \texttt{rns} \citep{1995Stergioulas}, is used to solve the relativistic stellar structure equations for axisymmetric rotating compact stars, assuming a zero temperature EOS.

\subsection{Strange Quark Star's EOS}\label{strEOS1}
For the strange quark star EOS two phenomenological EOS models have been implemented.
The first is the MIT bag model \citep{PhysRevD.9.3471,WEBER200794,PhysRevC.61.045203,Alford_2005}. For given values of the QCD coupling constant, $\alpha_c$, and the mass of the strange quark, $m_s$, \cite{Farhi:1984qu} have implemented a renormalization-group-improved QCD perturbative approach and found a range of possible values for the Bag constant, B, such that the bulk strange matter is stable. For instance, if $a_c=0.6$ and $m_s=100MeV$ we have $(145MeV)^4<B<(160MeV)^4$, while for $a_c=0$ and $m_s=200MeV$ we have $(130MeV)^4<B<(145MeV)^4$ \citep{Alcock:1988re}. In this work, 10 values have been chosen for the Bag constant, between 40 $MeV/fm^3$ and 100 $MeV/fm^3$ (or $(130MeV)^4<B<(167MeV)^4$).
 The pressure, P, is related to the energy density, $\epsilon$, as follows 

\begin{equation}
\begin{aligned}
P = \frac{1}{3}(\epsilon - 4B).
\label{Eqn:QS1}
\end{aligned}
\end{equation}

The second is the linear approximation introduced by \citep{2000A&A...359..311Z}. The pressure is related to the energy density as follows
\begin{equation}
\begin{aligned}
P = \frac{1}{3}(1+\epsilon_{fit})(\epsilon-\epsilon_s)
\end{aligned}
\end{equation}
with $\epsilon_s$ the surface density and $\epsilon_{fit}$ just a free parameter. Both can be estimated for given values of the bag constant, the QCD coupling constant, and the mass of the strange quark. Five such EOSs have been used with B = 60 $MeV/fm^3$ and ($m_s$,$\alpha_c$) = (50,0.1), (100,0.4), (150,0.3), (200,0), (250,0.6).

More realistic strange quark star EOSs exist in literature. For example, \cite{Issifu_2025} are using a modified density-dependent quark mass model, where a single-gluon interaction has been taken into account. The free parameters were calibrated to match the observational data \citep{2019Riley,Riley_2021,Doroshenko2022,Romani_2022}. \cite{2021PhRvD.103f3018Z} provide an EOS that takes into account the perturbative QCD correction and assumes color superconductivity for up, down and strange quarks. In this work, the former has been implemented to check the validity of the results.

\subsection{Inverse mapping}

One of the main outcomes of \citep{Konstantinou_2022} was that the gravitational mass ($M_*$) - radius ($R_*$) curve for non-rotating NSs can be reconstructed by the use of a set of empirical equations, and the gravitational mass ($M$), equatorial radius $R_e$, and spin frequency 
$\nu$ of rotating NSs. This is what is referred to as inverse mapping and is done as follows. 

First, the equatorial compactness of the rotating star is computed, defined as $C_e\equiv\frac{M}{R_e}$. Then $\Omega$ has been normalized as follows
\begin{equation}
    \Omega_{n2} = \frac{\Omega}{\Omega_{K2}},
\end{equation}
where $\Omega_{K2}$ is 

\begin{equation}
    \Omega_{K2} = \sqrt{\frac{GM}{R_e^3}} \times \sum_{i=0}^{4} b_i C_e^i.
    \label{eq:kepler-inv}
\end{equation}

Then the gravitational mass, $M_*$ and radius, $R_*$, of the non-rotating neutron star can be found by the use of 
\begin{equation}
\begin{aligned}
\frac{R_e}{R_*}^{nor.} \equiv  1 +  ( e^{A_{2} \Omega_{n2}^2} - 1 + B_{2} \left[ \ln( 1 - (\frac{\Omega_{n2}}{1.1})^4)\right]^2)
 \\ \times \left( 1 + \sum_{i=1}^{5} b_{r,i} C_e^i \right),
\end{aligned}
\label{eq:Rinvfit}
\end{equation}
and 
\begin{equation}
\frac{M}{M_*}^{nor.} \equiv 1 + \left( \sum_{i=1}^4 d_i \Omega_{n2}^i\right)  \times \left( \sum_{i=1}^{4} b_{m,i} C_e^i \right).
\label{eq:Minvfit}
\end{equation}
where the superscript ``nor." indicates that the relations apply to stars with normal EOS.
The coefficients $b_i$, $A_2$, $B_2$, $b_{r,2}$, and $b_{m,i}$ were empirically derived in order to describe the various EOSs, and can be found in Table~1 of \cite{Konstantinou_2022}.

In this paper, this inverse mapping process has been performed in order to check if the same universal-equations are valid for a self-bound system.
To verify the agreement of the universal relations with the produced data, the percent deviation has been computed. The percent deviation of a quantity $Q$ from its corresponding universal-relation predicted value $Q_{fit}$ is defined by
\begin{equation}
 Dev(Q) = 100 \times \frac{Q-Q_{fit}}{Q}.
\end{equation}

\begin{figure*}
\subfloat[]{\includegraphics[width = 3.5in]{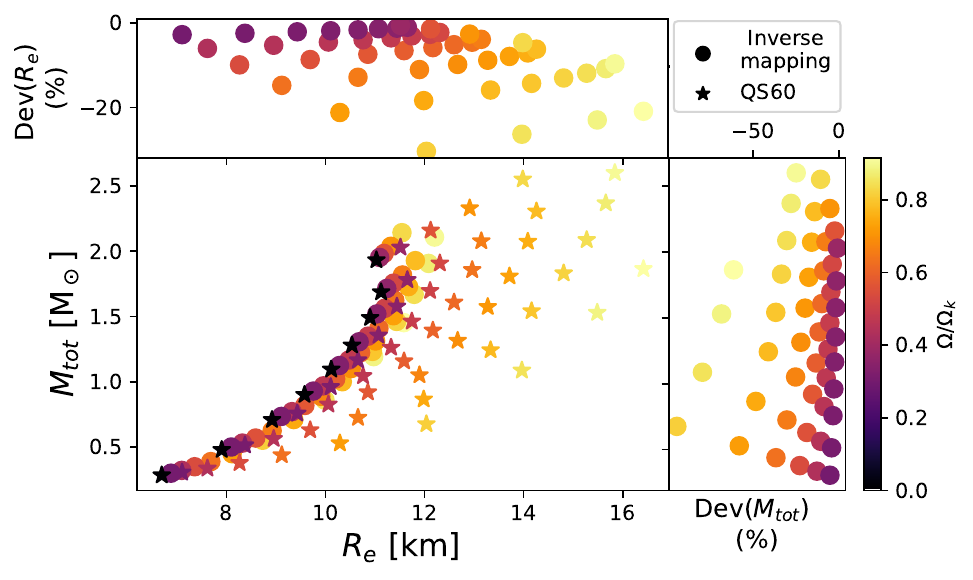}} 
\subfloat[]{\includegraphics[width = 3.5in]{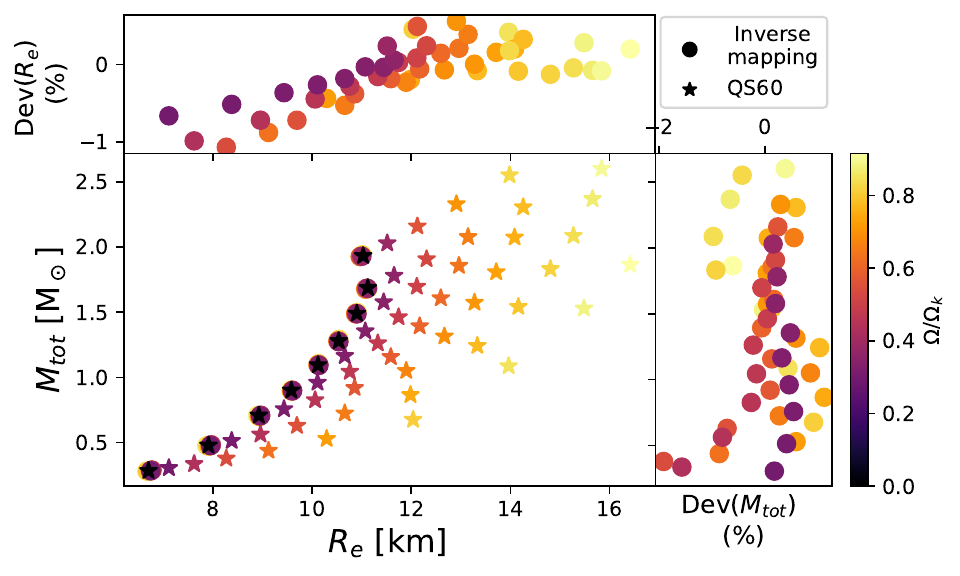}}\\
\subfloat[]{\includegraphics[width = 3.5in]{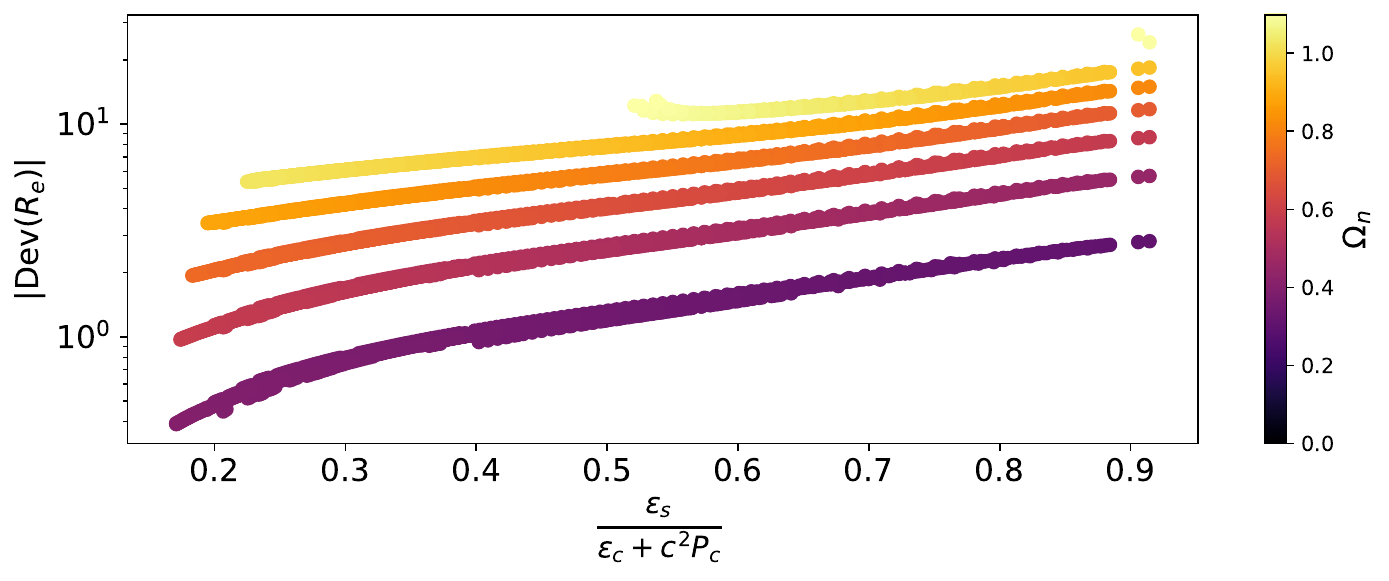}} 
\subfloat[]{\includegraphics[width = 3.5in]{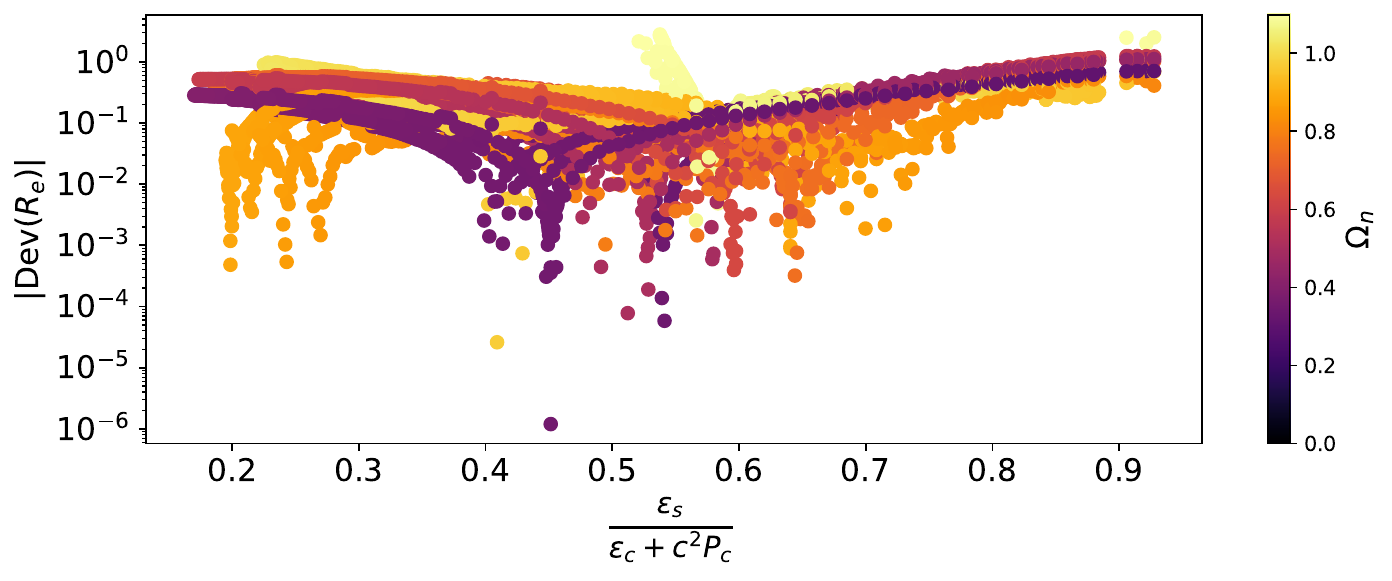}}\\
\subfloat[]{\includegraphics[width = 3.5in]{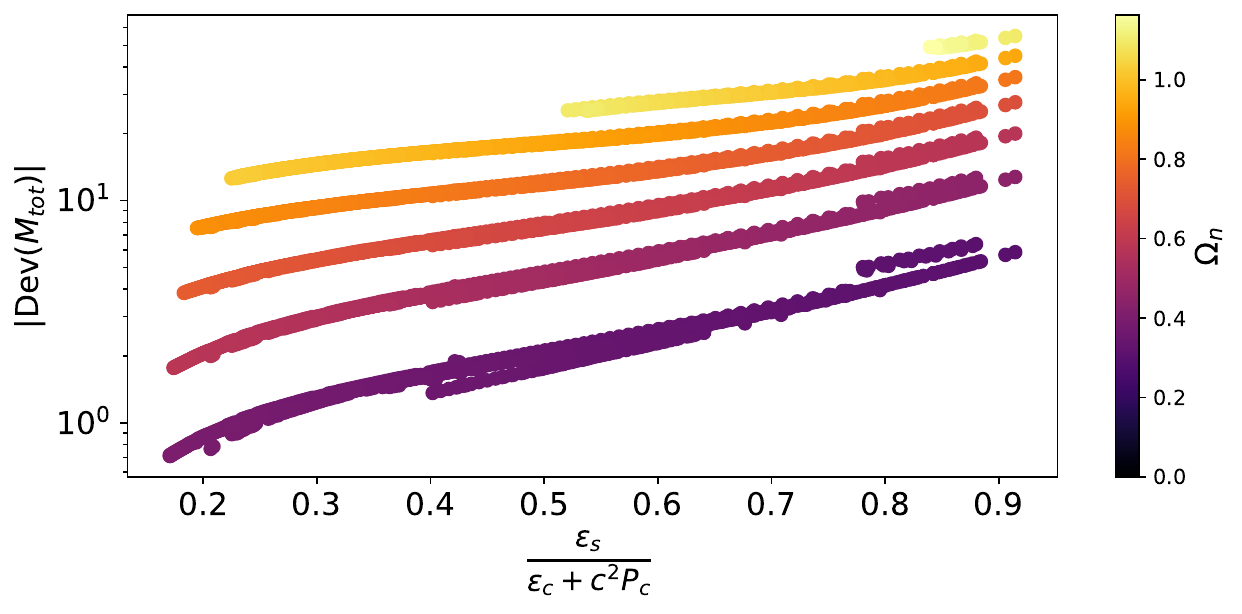}}
\subfloat[]{\includegraphics[width = 3.5in]{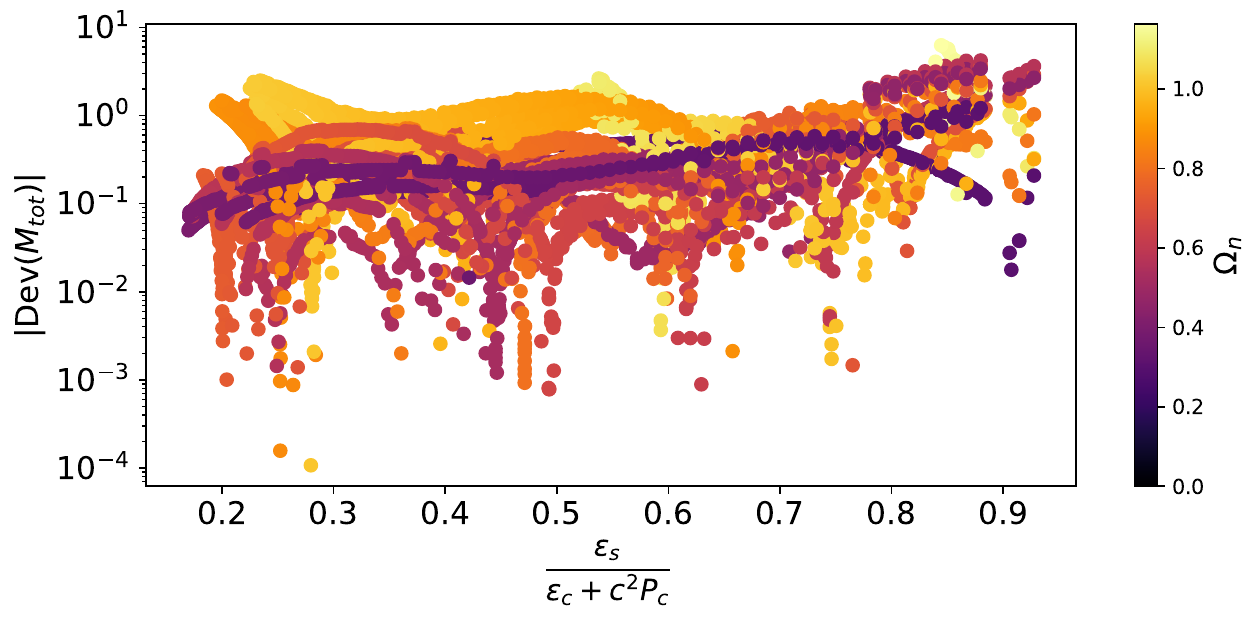}} 
\caption{Mass-radius plots for rotating strange quark stars. The star points in the mass-radius plots (a) and (b), represent strange quark stars with B = 60 $MeV/fm^3$ (QS60), with their corresponding spin frequencies-to-Kepler frequency ratio indicated by the color gradient. The dotted colored points illustrate the results of inverse mapping, as predicted by the universal relations. The difference in (a) and (b) is that for the inverse mapping in (a) the old universal relations has been used, while for (b) the updated version of the universal relations is adapted. Above and to the right of figures (a) and (b), the percent deviation of the equatorial radius and gravitational mass from the universal relations derived for normal EOSs is displayed. Figures (c) and (d) show the percent deviation of the change of the equatorial radius, while (e) and (f) show the percent deviation of the change of the gravitational mass, of the 15 strange quark star EOSs created in Section \ref{strEOS1}. Again, the plots on the left show the case where the old universal relations have been used, while the plots on the right illustrate the update that is introduced in this work. This visual representation allows for a clear comparison between the 15 self-bound stars and the predictions from normal EOS-based universal relations.}
\label{fig:results}
\end{figure*}

\section{Results}\label{sec:Results}
\subsection{Improved Universal Relations}

In this section, it has been investigated whether the gravitational mass and radius universal relations hold for strange quark star EOSs. To test this, the 15 strange quark star EOSs generated in Section \ref{strEOS1} were used. 9015 stars have been computed using the \texttt{rns} code, with gravitational masses $0.13\ M_\odot<M<4.9\ M_\odot$ and equatorial radii $4.5\ km<R_e<31\ km$. The gravitational masses and radii of rotating strange quark stars modeled with the choice of B = 60 $MeV/fm^3$ and by the use of Equation \ref{Eqn:QS1}, referred to as QS60, are illustrated in Figure \ref{fig:results} (a), where the color-map shows the star's spin to Kepler frequency ratio. Only a subset of the data has been plotted in order to make this figure easier to read.

The same figure also shows the results of the inverse mapping, where the properties of rotating strange quark stars are used to reconstruct the non-rotating limit. The plots above and adjacent to the gravitational mass-radius diagrams indicate the percent deviation of the equatorial radius and the gravitational mass of QS60 from the expected values based on universal relations. Figures \ref{fig:results} (c) and (e) show these deviations versus $\frac{\epsilon_s}{\epsilon_c+c^2P_c}$, and the normalized frequency, $\Omega_{n2}$, for all the 15 EOS. Here $\epsilon_s$ and $\epsilon_c$ are the surface and central energy density, respectively, and $P_c$ is the pressure at the center.

It can be seen that as $\frac{\epsilon_s}{\epsilon_c+c^2P_c}$ and frequency increase, the deviation from universal relations also increases. The maximum deviations reach approximately 25\% in radius and 55\% in gravitational mass. Therefore, it can be concluded that the universal relations proposed by \cite{Konstantinou_2022} for gravitational mass and radius changes do not hold for strange quark star EOSs, which represent the physically relevant class of self-bound configurations.

To reduce the deviations caused by the strange quark star EOS, new empirical relations are proposed. These revised relations aim to better capture the effects of self-bound systems. The data are fitted by using the following equations
\begin{equation}
\begin{aligned}
\frac{R_e}{R_*}^{s.b.} \equiv  \frac{R_e}{R_*}^{nor.}\times \sqrt{\frac{|1-\frac{\epsilon_s}{\epsilon_c+c^2P_c}\times C^R|}{|1-\frac{\epsilon_s}{\epsilon_c+c^2P_c}\times C^R\times (1+F^R) |}},
\end{aligned}
\label{eq:Rinvfitnew}
\end{equation}
and 
\begin{equation}
\frac{M}{M_*}^{s.b.} \equiv \frac{M}{M_*}^{nor.}\times (1+D^M \times C^M \times F^M).
\label{eq:Minvfitnew}
\end{equation}
Here, the superscript ``s.b." indicates that the relations only apply to self-bound stars. The empirical functions that are introduced above are defined as 
\onecolumngrid

\begin{table}[t]
\caption{The $i^{th}$ component of the $c^R_i$, $f^R_i$, $c^M_i$, $f^M_i$ and $d^M_i$ fitting coefficients, and the $R^2$ statistic  for the best-fit equations. }
    \label{tab:coeff}
    \begin{tabular}{|cc|c|c|c|c|c|c|c|}
    \hline
Equation & $R^2$ &  Symbol & $i=0$ & $i=1$ & $i=2$ & $i=3$ & $i=4$ & $i=5$  \\
\hline
 (\ref{eq:Rinvfitnew}) & 0.9993& $c_i^R$ & -48.2636 & 0.03748 & -0.9810& - & - & -\\
 && $f_i^R$ & - & -1.7979 &  15.1950 & - & - & -\\
\hline
 (\ref{eq:Minvfitnew}) & 0.9989& $c_i^M$ & - & 5.6402 & -75.5032 & 360.96 & -551.20 & -\\
 && $f_i^M$ & - &  6.8264 & -13.7904 & 32.0368 & 8.4020  & -\\
  && $d_i^M$ & - & 1.1813 & -5.3497 & 14.0525 & -17.9481 & 8.9275\\
\hline
    \end{tabular}
\end{table}

\twocolumngrid
\begin{equation}   
\begin{aligned}
    D^M=\sum_{i=1}^{5} d^M_i (\frac{\epsilon_s}{\epsilon_c+c^2P_c})^{2i},\ \ \ \ \ \ \ 
    \\C^M=\sum_{i=1}^{4} c^M_{i} C_e^{i},\ \ \ \ \ \ \ \
    F^M=\sum_{i=1}^{4} f^M_i \Omega_{n2}^{i},
    \\C^R=\sum_{i=0}^{2} c^R_i C_e^i, \ \ \ \ \ \ \ \ \ F^R=\sum_{i=1}^{2} f^R_i \Omega_{n2}^i,
\end{aligned}
\end{equation}
and depend on the fitted parameters$c^R_i$, $f^R_i$, $c^M_i$, $f^M_i$ and $d^M_i$, which can be found in Table \ref{tab:coeff}.

The choice of this specific form for the corrections will be explained in Section \ref{sec:Discussion}. Figures \ref{fig:results} (b), (d), and (f) is the re-creation of Figure \ref{fig:results} (a), (c), and (e), but by using the new empirical relations \ref{eq:Rinvfitnew} and \ref{eq:Minvfitnew} for the inverse mapping. The improved fits reduce the maximum deviation of the equatorial radius to approximately 2.8\% and the gravitational mass deviation to around 6.7\%, significantly enhancing the accuracy of the universal relations for strange quark star EOSs. Similar deviations characterize the stars where the EOS by \cite{Issifu_2025} was used (with 4\% and 8\% the maximum deviations, respectively).

\section{Understanding Deviations from Universal Relations}\label{sec:Discussion}

To interpret the significant deviations exhibited by self-bound EOSs from the universal relations presented in Section \ref{sec:Results}, the physical origin of these relations is discussed in the following.

\subsection{Conservation of Gravitational Potential Differences in Rotating General Relativistic Stars}

In the general relativistic case, a time-independent axisymmetric metric is adopted to describe rotating stars \citep{1971ApJ...167..359B}:
\begin{equation}
\begin{aligned}
    ds^2 = -e^{2\Phi^{GR}} dt^2 + e^{2\zeta - 2\Phi^{GR}}(dr^2 + r^2 d\theta^2) \\+ B^2 e^{-2\Phi^{GR}} r^2 \sin^2\theta (d\phi - \omega dt)^2,
    \end{aligned}
\end{equation}
where $\Phi^{GR}$, $\zeta$, $B$, and $\omega$ are metric functions depending on $r$ and $\theta$.

The stationarity (axisymmetry) of the metric, $\frac{\partial g_{\mu \nu}}{\partial t}=0$ ($\frac{\partial g_{\mu \nu}}{\partial \phi}=0$), will introduce the $\xi_\mu$($\chi_\mu$) killing vector satisfying the Killing equation $\nabla_\mu \xi_\nu(\chi_\nu)+\nabla_\nu \xi_\mu(\chi_\mu)=0$. Combining this with the assumption of rigid rotation, and a perfect fluid energy-momentum tensor, Bernoulli's theorem yields (see Section 3.4 of \cite{gourgoulhon2011introduction})
\begin{equation}
N(P)+\Phi^{GR}+ln(1-v^2)/2=\text{constant}.
\end{equation}
where $v = (\Omega - \omega) r \sin\theta B e^{-2\Phi^{GR}}$, N(P) is relativistic specific enthalpy given by
\begin{equation}
    N(P)-N_p = \int_{P_p}^P \frac{dP}{\epsilon + P},
\end{equation}
where $\epsilon$ is the energy density, and the subscript $p$ denotes that the quantity is evaluated at the pole of the star.

Since for rigid rotation $v$ is zero at the center and at the poles, it is easy to show that
\begin{equation}
    \Phi^{{GR}}_p - \Phi^{{GR}}_c = N_c-N_{p}.
\end{equation}

Thus, for sequences of rotating compact stars sharing the same boundary conditions (i.e. fixed $N_c$ and $N_p$), which are the ones considered in this work, the difference in gravitational potential between center and pole remains constant along the sequence
\begin{equation}
    \Phi^{{GR}}_c - \Phi^{{GR}}_p = \text{constant along a sequence}.
\end{equation}

\subsection{Conservation of Gravitational Potential Differences in Rotating Newtonian Stars}
For simplicity, consider the Newtonian case. For a rotating, axisymmetric star in Newtonian gravity under hydrostatic equilibrium, the force balance condition reads
\begin{equation}
    -\nabla P(r,\mu) = \rho(r,\mu)\nabla \left[\Phi(r,\mu) + \Psi(r,\mu)\right],
\end{equation}
where $r$ is the radial distance and $\mu = \cos\theta$ is the cosine of the polar angle. Here, $P$ is the pressure, $\rho$ is the mass density, $\Phi$ is the gravitational potential, and $\Psi$ is the centrifugal potential, which depends on the rotation profile. For rigid rotation, the centrifugal potential takes the form $\Psi(r,\mu) = -\frac{1}{2}\Omega^2 r^2 (1 - \mu^2)$.

Imposing the condition that $\Psi$ vanishes at the center and along the polar axis (which means $\Psi(0,\mu) = \Psi(r,1) = 0$), and defining the differential specific enthalpy as $dh = \frac{dP}{\rho}$, the force balance equation can be integrated to produce the following
\begin{equation}
    h(P(r,\mu)) + \Phi(r,\mu) + \Psi(r,\mu) = \text{constant}.
\end{equation}

Evaluating this expression at the pole ($\mu = 1$) and center ($r = 0$), where $\Psi = 0$, we get
\begin{align}
    h_c - h_p + \Phi_c - \Phi_p = 0.
    \label{eq.hyd}
\end{align}
Again, for sequences where $h_c$ and $h_p$ are fixed, the difference in gravitational potential between center and pole remains constant
\begin{equation}
    \Phi_c - \Phi_p = \text{constant along a sequence}.
\end{equation}

Since, $\Phi \sim M/R_e$ we expect that $M/R_e$, remains nearly constant along a sequence \citep{Konstantinou_2022}.
\subsection{Radius Change Universality from the Conservation of Gravitational Potential Differences}
The Newtonian potential is defined at distance r and angle $\theta$ as \citep{1986Hachisu}
\begin{equation}
\begin{aligned}
\Phi(s,\mu)=- 2\pi G \rho_c R_e^2 \int_{s}^{1}ds' \int_{-1}^{1}d\mu' \\\sum_{n=0}^{\infty} P_{2n}(\mu)P_{2n}(\mu') \tilde{\rho}(s',\mu') \frac{f_{2n}(s',s)}{(1-s')^2}
\label{pot}
\end{aligned}
\end{equation}
where $r/R_e=s/(1-s)$ and $\mu=cos\theta$, and $f_{2n}(s',s) = (\frac{s}{1-s})^{2n+1} (\frac{s'}{1-s'})^{2n+2}$ when $s'<s$, and $f_{2n}(s',s) = (\frac{s}{1-s})^{2n}(\frac{1-s'}{s'})^{2n-1}$ when $s'>s$. The scaled mass density profile $\tilde{\rho}(s',\mu')$ is defined as the density divided by the central density $\rho_c$, such that $0<\tilde{\rho}(s',\mu')<1$.

The equatorial radius can be obtained from Eqn. \ref{eq.hyd}
\begin{equation}
\begin{aligned}
 R_e^{-2} = \frac{\rho_c}{h_c -h_p}\Delta \tilde{\Phi}.
\end{aligned}
\end{equation}
where $\Delta \tilde{\Phi}\equiv \tilde{\Phi}_p -\tilde{\Phi}_c$, with $\tilde{\Phi}\equiv  \frac{\Phi}{\rho_c R^2_e }$.

The change of the equatorial radius between two stars with different equations of state and spin frequency will be 
\begin{equation}
\begin{aligned}
 \frac{R_e}{R_*} = \sqrt{\frac{\rho^*_c}{\rho_c}\frac{(h_c -h_p)}{(h^*_c -h_p^*)}\frac{\Delta \tilde{\Phi}^*}{\Delta \tilde{\Phi}}},
 \label{eq:rad-ch}
\end{aligned}
\end{equation}
where the asterisk has been used for the non-rotating star's microscopic parameters and $R_*$ for its equatorial radius, while non-starred parameters and $R_e$ were implemented for the rotating case.

Here, we are mainly interested in the study of the change in the equatorial radius while we have the same EOS and boundary conditions ($h^*_c=h_c$, $h^*_p=h_p$, $\rho^*_c=\rho_c$, $\rho^*_s=\rho_s$). 

Sequences with constant baryon mass are appropriate for modeling the spin down of isolated compact stars \citep{1994Cook}. But in this case $\frac{R_e}{R_*}$ will depend on both $\frac{\rho^*_c}{\rho_c}$ and $\frac{h_c -h_p}{h^*_c -h^*_p}$, as the central and surface density and enthalpy must change along the sequence for the baryon mass to remain constant. Thereafter, the change of the equatorial radius in such sequences will depend on the choice of the EOS. This is the reason why we investigate a sequence in a parameter space where $\frac{\rho^*_c}{\rho_c}=\frac{h_c -h_p}{h^*_c -h^*_p}=1$, simply because this eliminates the EOS dependence coming from $ \frac{R_e}{R_*} \propto \sqrt{\frac{\rho^*_c}{\rho_c}\frac{h_c -h_p}{h^*_c -h^*_p}}$. Such sequences are also implemented by EOS inference codes \citep{PhysRevD.108.124056}.

We could approach the density profile of a normal star, $\rho^{nor.}(r,\theta)$, as a combination of a self-bound density profile, $\rho^{s.b.}(r,\theta)$ and a crust $\rho^{cr}(r,\theta)$,
\begin{equation}
\begin{aligned}
\rho^{nor.}(r,\theta)\approx\begin{cases} 
\rho^{s.b.}(r,\theta)   &,\ \rho>\rho_s, \\\rho^{cr}(r,\theta)
 &,\ \rho<\rho_s.
\end{cases}\ \ \ 
\label{eqn:dens-norm}
\end{aligned}
\end{equation} 
where $\rho_s$ is the surface density value of a self-bound profile. 

The linearity of the Newtonian gravitational field equations leads to the result that the gravitational potential of a star with normal density, $\Phi^{nor.}$, will be related to the one with self-bound one, $\Phi^{s.b.}$ as
\begin{equation}
\begin{aligned}
\Phi^{s.b.}=\Phi^{nor.}-\Phi^{cr},
\end{aligned}
\end{equation} 
where $\Phi^{cr}$ is the contribution to the total gravitational potential only from the crust alone. By adapting $\rho^{s.b.}=\rho_c(\tilde{\rho}^{nor.}\Theta(\rho-\rho_s)-\frac{\rho_s}{\rho_c}\tilde{\rho}^{cr}\Theta(\rho_s-\rho))$ into Eqn. \ref{pot}, where $\Theta$ is a Heaviside function, it is not difficult to show that 
\begin{equation}
\begin{aligned}
\tilde{\Phi}^{s.b.}=\tilde{\Phi}^{nor.}-\frac{\rho_s}{\rho_c}\tilde{\Phi}^{cr}.
\end{aligned}
\end{equation} 

Applying this density decomposition in Eqn. \ref{eq:rad-ch}, we arrive at a generalized expression for the radius change in terms of the potential difference and the surface-to-central density ratio, defined as
\begin{equation}
\begin{aligned}
\label{eqn:chnorm}
\frac{R_e}{R_*} ^{s.b.}
= \frac{R_e}{R_*}^{nor.} \sqrt{\frac{1-\frac{\rho_s}{\rho_c}\Delta \tilde{\Phi}^{*cr}/\Delta \tilde{\Phi}^{*nor.}}{1-\frac{\rho_s}{\rho_c}\Delta \tilde{\Phi}^{cr}/\Delta \tilde{\Phi}^{nor.}}},
\end{aligned}
\end{equation}
\\
where $\frac{R_e}{R_*}^{nor.}$ is the change of the equatorial radius of a star with a normal density distribution, and is equal to $\sqrt{\frac{\Delta \tilde{\Phi}^{nor.}}{\Delta\tilde{\Phi}'^{nor.}}}$. We expect that when $\frac{\rho_s}{\rho_c}\Delta \tilde{\Phi}^{*cr}/\Delta \tilde{\Phi}^{*nor.} \text{ and }\frac{\rho_s}{\rho_c} \Delta \tilde{\Phi}^{cr}/\Delta \tilde{\Phi}^{nor.}$ $<<1$, and at the slow rotation limit, where $\Delta \tilde{\Phi}^{*cr}/\Delta \tilde{\Phi}^{*nor.} \approx \Delta \tilde{\Phi}^{cr}/\Delta \tilde{\Phi}^{nor.}$ that $\frac{R_e}{R_*}^{s.b.}\approx \frac{R_e}{R_*}^{nor.}$, and therefore universality will hold for self bound EOSs too. This explanation agrees with what has been seen in Figure \ref{fig:results}.

Equation \ref{eqn:chnorm} reveals how the deviation scales with the surface-to-central density ratio. This motivates the correction factors in Eqn. \ref{eq:Rinvfitnew}, which incorporate $\frac{\epsilon_s}{\epsilon_c+c^2P_c}$ (the relativistic analogue of $\frac{\rho_s}{\rho_c}$) to improve inverse mapping.

\section{Conclusion}
\label{sec:conclusion}
In this paper, the impact of a strange quark star equation of state on the structure and rotational properties of compact stars is investigated, with a particular focus on deviations from established universal relations. By examining sequences of strange quark stars with constant central energy density, significant deviations has been identified in the gravitational mass-radius relationship, especially for stars with high surface-to-central total energy density ratios.

Our analysis shows that the gravitational potential difference between the center and poles, as well as the metric potentials in general relativistic stars, remain constant within sequences of rotating stars. This conservation leads to a quasi-universality in the equatorial radius's response to rotation, explaining some of the observed deviations from expected universal relations.

New empirical relations have been introduced that better capture the behavior of self-bound strange quark stars, and verified their accuracy through inverse mapping. These improved relations offer a significant reduction in the deviation of both gravitational mass and radius predictions, making them useful for modeling rapidly rotating compact stars with self-bound EOSs. 

For present NICER and gravitational wave applications to slowly rotating objects, the rotation corrections derived here are typically smaller than current statistical and systematic uncertainties and are therefore not essential in standard EOS-inference workflows. In future, when measurements improve and precise radii for faster pulsars become available, these new universal relations can be incorporated into EOS inference code to sharpen tests of the strange star hypothesis.

\section{Acknowledgements}
I acknowledge financial support from “The three-dimensional structure of the nucleon from lattice
QCD” 3D-N-LQCD program, funded by the University of Cyprus, and from the projects HyperON (VISION ERC - PATH 2/0524/0001) and Baryon8 (POST-DOC/0524/0001), co-financed by the European Regional Development Fund and the Republic of Cyprus through the Research and Innovation Foundation. I would like to express my gratitude to Sharon Morsink for valuable comments, suggestions, and discussions on this paper, and for her support over the years. Furthermore, I would like to thank Adamu Issifu for our insightful discussion and for sharing his strange quark star EOSs for further validation of the results of this paper. Additionally, I extend my thanks to Demetrianos Gabriel, Anastasios Eracleous and Evelyn Toumazou for our insightful conversations. Laslty, I would like to thanks Marinos Vlasis, Kwstas Festas and Andreas Ksistras for their comments on the structure of the paper.

\bibliography{biblio}{}
\bibliographystyle{aasjournal}

\listofchanges

\end{document}